# PHENOMENOLOGICAL MESOSCOPIC MODELS FOR SEIZURE ACTIVITY

Maria Luisa Saggio and Viktor K. Jirsa

*Institut de Neurosciences des Systèmes, UMR Inserm 1106, Aix-Marseille Université, 9 Faculté de Médecine, 27, Boulevard Jean Moulin, 13005 Marseille, France*

marisa.saggio@gmail.com
viktor.jirsa@univ-amu.fr

## Abstract

In this chapter we review phenomenological models of seizure like activity. We discuss dynamical mechanisms for seizure onset and offset, preictal spikes, spike and wave complexes and status epilepticus, highlighting the role played by the bifurcation structure of the model, the presence of noise and the emergence of multiple interacting time-scales. These models can be used to build large-scale patient specific brain network models serving as in-silico platforms to test clinical hypothesis and perform virtual surgeries. They suggest innovative treatment strategies, such as minimally invasive ablations or stimulations that fully exploit the network and dynamical properties of the system, or even modulation of variables and parameters to force the system in safer regions of the bifurcation diagram. We discuss insights from phenomenological models that can help to foster our understanding of the mechanisms underlying epileptic seizures.

## 1 Introduction

Epilepsy is the most common among the chronic and severe neurological diseases, affecting 65 million people worldwide and is characterized by an augmented susceptibility to seizures. Seizures are "transient occurrence of signs and/or symptoms due to abnormal excessive or synchronous neuronal activity in the brain." [1]. Current therapeutic strategies have the goal of suppressing or reducing the occurrence of seizures, being thus symptomatic rather than curative, and there are not known therapies able to modify the evolution of acquired epilepsy, or to prevent the development of this. Furthermore, 25-40% of patients do not respond to pharmacological treatment, and this number stays unchanged when using new generation anti-epileptic drugs as compared to established ones. For drug-resistant patients with focal epilepsy (an epilepsy in which seizures start in one hemisphere) there exists an alternative to medication, that is surgical resection of the brain regions involved in the generation of seizures, the epileptogenic zone (EZ), under the constraints of limiting post-surgical neurological impairments. Rates of success of brain surgery for epilepsy treatment vary between 34% and 74% as a function of the type of epilepsy. Outcomes are very variable, depend on the patient condition and epilepsy and can change in time.

Focal epilepsies involve widespread networks, so they are considered a network disorder [2]. From a network perspective, seizures depend on the interaction between the excitability of the single nodes of the network and the connections among them. What defines a node depends on the spatial scale of the problem we are dealing with, so that nodes can be neurons at the microscopic scale or brain regions at the mesoscopic one. While we have a better understanding of phenomena acting at the neuronal scale rather than mesoscopic one, clinicians have to base their hypothesis and decisions on mesoscopic observables when planning surgical strategies for drug-resistant patients. Efforts towards the understanding of mechanisms of seizure generation and propagation at this scale could provide clinicians with additional tools in their evaluation.

The spatial component is not enough to understand seizures: epilepsy is also a dynamic disease [3] and seizures exhibit a complex temporal evolution, as observed for example through electroencephalography (EEG) or Stereotactic EEG (SEEG) recordings. These measurements reflect the global electric activity generated by a huge number of neurons. While the degree of spike synchronization among neurons has a heterogeneous, highly variable and complex pattern, seizures are characterized by synchronized activity at the level of EEG and local field potential, especially during seizure evolution and termination. This implies a decrease in the degrees of freedom necessary to describe the activity of the underlying system, which may be amenable to mathematical modeling through the use of mesoscopic collective variables for the activity of a single brain region (i.e. a node in the network).

Large-scale mathematical and computational models have the potential to merge networks and dynamics to generate non-trivial predictions, which can push our understanding of the meso- and macro-scopic mechanisms of seizure generation and evolution. In addition they can provide a platform to perform in silico experiments, such as testing hypothesis on where the seizure starts (EZ) and where it propagates (Propagation Zone, PZ), perform systematic virtual resections to establish the best surgical strategy, as well as proposing new approaches to prevent seizures by manipulations of quantities represented by other parameters of the network or of the dynamics [4]–[8]. Models can be informed with detailed patient-specific information, in terms of large-scale connectivity, presence of lesions and functional information, on the line of personalized medicine [5].

At the level of a single node, there are different approaches to model the activity of a brain region. One possibility is to create a model that attempts to mimic physiologically realistic mechanisms and variables. The large number and non-identifiability of parameters poses one of the big challenges in model inference and the estimation of correlated parameters from empirical data is notoriously difficult. Errors, even small ones, in the values of these parameters, the fact that some parameters' values may be context dependent (e.g. they may depend on the specific patient) or that they can evolve in time, may dramatically alter the dynamics of the model. While physiological models can foster our understanding of such specific mechanisms, they are often computationally expensive and may not be well suited to perform extensive simulations and parameter sweeps. Complementary approaches focus on the development of lower-dimensional models aiming at a faithful reproduction of data features, often rooted in principles derived from statistics or mathematics and linked to data structure rather than physiology. Such a *phenomenological* approach decreases the computational cost of the model and renders it a prime candidate for routine investigations in the clinics. Moreover, this focus on the essential dynamics provides a more tractable model amenable to mathematical investigation of its properties. Investigations that rely on realism in terms of dynamics, such as the study of the pattern of seizure propagation or of the reaction of a brain region to stimulation, can thus benefit from the use of phenomenological models. Furthermore, while phenomenological variables lack a clear correlate with the physical substrate, they allow to model observed phenomena for which the underlying physiology is not sufficiently clear or for which multiple physiological mechanisms may produce the same outcome, as for epileptic seizures. Despite the large range of seizure mechanisms acting on different time and spatial scales, electrographic signatures of these events are relatively stereotyped across different pathologies, and even among different species and primitive laboratory models. Seizures can also be induced in any otherwise healthy brain, from flies to humans and both in vivo and in vitro, and the electrographic signatures of these induced seizures are similar among them despite the variety of provoking conditions which may be used [9]. Such similarity among seizures points towards the existence of invariant dynamical properties in their underlying mechanisms across brain scales, regions and across species [9] and phenomenological models are ideal to capture and study such key invariant properties.

In this chapter we will review meso- and macro-scopic phenomenological models for seizure activity. We will start with a brief introduction to how the brain can be conceptualized as a complex system and how this allows to build large-scale brain models. We will then show how this framework can be applied to the study of epilepsy, focusing on the description of mesoscopic phenomenological models for a brain region able to generate seizures. To conclude, we will discuss applications and insights coming from these models.

## 1.1 Network approach to the brain

Complex systems are ubiquitous in our universe. From the macroscopic world of galaxies and stars to the microscopic systems of cells, molecules and atoms and even further. The surprising fact is that heterogeneous complex systems often show similar organizational properties when analyzed as networks. To build a network out of a complex system we need to identify two kinds of entities: nodes, which are the elements of the network, and links, which express some kind of connection between those elements, such as physical wiring or functional relationship. The same complex system can be conceptualized as a network in many different ways: the particular choice of which elements constitute the nodes and of which characteristics of their connections entail a link will depend on the specific goals of the investigation undertaken. Links can be summarized in the connectivity matrix, a squared matrix having the same dimension as the number of nodes in the system and whose entries express the pairwise value of the link between any two nodes.

The brain can certainly figure among the most complex system: $\sim 10^{11}$ neurons and $\sim 10^{14}$ synaptic connections among them, not even to count the further degrees of freedom within each single neuron (dendritic branching, neurotransmitters and so on) or the fact that the brain is constituted by other cell types other than neurons, such as glia, which greatly outnumber them. Despite this astonishing amount of elements and the intricacies of the wiring, the brain exhibits an impressive level of organization and network approaches are well suited to investigate these characteristics.

At the microscale, the building block of our brain is the neuron, at least for the purposes of this chapter. Molecular dynamics, especially in synapses, plays an increasingly important role in translational neuroscience and drug design [10] but shall not be further reviewed here. We can describe a neural network considering each neuron as a node and each synaptic connection between a pair of them as a link. In addition, we need to consider the dynamics of the single units and of the couplings to describe the temporal evolution of the network. The number of neurons that can be simulated with current technology depends on the level of details of the model used ranging from a million of neurons with detailed morphology (Human Brain Project (http://www.humanbrainproject.eu) to 500 million of highly simplified neurons (for instance, DARPA Synapse project) [11]. These simulations, even with simple neurons, require an enormous computational power and are currently not feasible to be performed routinely and in a personalized fashion in clinical settings. In addition, while such detailed microscopic simulations contribute to our understanding of some physiological brain mechanisms, there are questions that need a different, mesoscopic level of description. This level allows a comparison between simulation output and measures from non-invasive functional brain imaging techniques, such as EEG, functional Magnetic Resonance Imaging (fMRI) or Magnetoencephalography (MEG), as well as intracranial recordings such as SEEG used in the clinical exploration of epilepsy. A trade-off between large coverage and microscopic details can be reached through a mesoscopic level of description, able to reduce the degrees of freedom of a whole brain region to just a few variables [12]. Such a representation is more tractable and less computationally expensive. In addition, it puts emphasis on behaviors that may emerge at the mesoscopic scale and are not present in the microscopic description.

In large-scale brain models, the brain is usually parcellated in several mesoscopic cortical and subcortical regions connected among them. While the first large-scale network brain models used homogeneous connectivities, developments in diffusion imaging techniques of the last two decades provide subject specific information on the brain's large scale connectivity established by the white matter and may provide more sophisticated and personalized network constraints. Diffusion imaging is an MRI based technique that tracks the diffusion of water molecules, which in the brain is maximally oriented along the axonal fibers. These data can be used to build a structural connectivity matrix, called connectome, reflecting the presence and weight of anatomical connections. The connectome has recently been extended to include also the matrix of estimated tract lengths, which determine the time delays of signal propagation due to finite velocity along white matter fibers, as this can strongly affect the network behavior. The connectome has been used as a basis to build large-scale brain models for resting state activity and to understand the role of noise and time delays in the generation of resting state fluctuations [13]. Since then the use of connectomes in network models has been followed up by many studies and connectome-based modeling has become a mature field.

Brain regions serve as network nodes in connectome-based brain models. At the level of a single node, the dynamics of a brain region is described using a small number of collective variables accounting for the global activity of the underlying networks of neurons. There exist formal approaches to perform this huge reduction of degrees of freedom, such as mean-field approximations, that benefit from results of statistical physics. Such approaches are widely used in physics. For example, under specific assumptions, we can describe the behavior of particles in a gas using collective variables such as temperature or pressure. In a brain region, neurons having similar or identical statistical properties and receiving similar or identical inputs are grouped into neural populations. The activity of each population is described by a probability density function for relevant neuronal states (e.g. the firing rate). In the presence of strong coherence among the neural states within a population, it is possible to restrict the model to the mean of the distribution, further reducing the degrees of freedom of the problem and allowing the study of several interacting populations within a single brain region. This provides the so-called Neural Mass Model (NMM) [13]. The dynamics of a neural population can mirror that of each single neuron composing it, or the NMM can be informed using empirical observations of how a system responds to its inputs, to account for the fact that different levels of organization of complex system (micro- vs meso- or macro-scopic) may obey different rules [14]. A well-known example of the latter approach is the Wilson-Cowan model, which includes an excitatory population of pyramidal neurons interacting with an inhibitory population of interneurons.

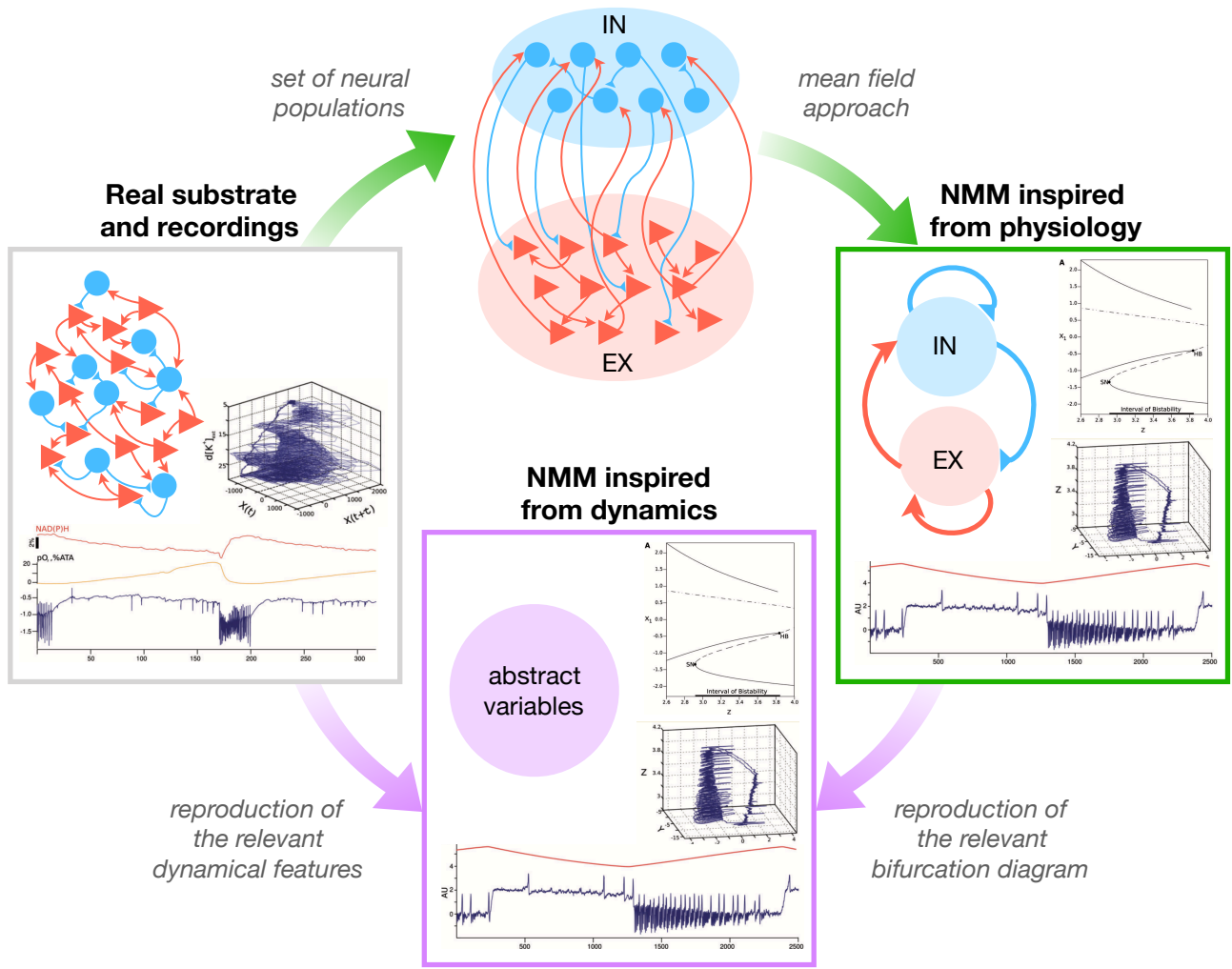

*Figure 1: Building Neural Mass Models (NMMs). NMMs reproduce the mesoscopic global activity of underlying networks of neurons. Common steps for the construction of physiologically inspired NMMs (green arrows) include: the identification of clusters of neurons sharing similar statistical properties and connectivity - neural populations; the application of mean field reduction techniques to obtain the dynamics of the collective variables describing each neural population and connections among them. In the figure we have schematically represented only two types of neurons (and thus neural populations), excitatory (red) and inhibitory (blue) ones, but NMMs can include several types. The model obtained can be used to reproduce the observed time series. In addition, one could study the role of parameters and produce bifurcation diagrams to describe the dynamical repertoire of the model. Phenomenological models (lavender arrows), can be obtained in two ways: directly from data by identifying relevant features and reproducing them; or by building simpler models with an equivalent bifurcation diagram to that of physiological NMMs. Phenomenological models inspired by physiological ones do not aim at reproducing the full dynamical repertoire of the latter, but a subset of the dynamics which is judged relevant for a certain scope.*

Typically, NMMs are inspired from physiology, comprise different neural populations and exhibit non-linear dynamics. As mentioned before, another possibility is to model specific dynamical properties of the brain region, without attempting a realistic physiological implementation (figure 1). In some cases, a direct link can be drawn among the variables of these phenomenological models and neural populations activities [15]–[17], so that the separation between NMMs and phenomenological ones is fuzzy. For this reason, we will refer to any model for the global activity of a brain region as a NMM, and will distinguish whether it is inspired from the neurophysiology or whether the model is phenomenological, i.e. inspired from the dynamics.

NMMs use differential equations to model the activity of a single node of the network and lack a spatial component. This limitation can be overcome, for example, by the use of Neural Field Models (NFMs). They can produce different spatiotemporal patterns of brain activity, including traveling waves, which seem to have an important role in the spreading of the ictal wave-front on the cortical surface [18], [19].

Connectivity, time delays and node dynamics can be combined, together with the addition of the dynamics of the couplings between nodes, in a large-scale brain model. A generic mathematical formulation for such a computational model, referred to as Brain Network Model (BNM) [15], or also graph-based brain anatomical network [20], includes other elements, such as the local connectivity within a brain region. The dynamics of the full BNM can be implemented with limited computational resources, provided that the number of nodes in the parcellation is not too high. Many studies use parcellations with order 100 regions, even though smaller or larger parcellations are possible (see [21] for the effect of parcellation size on simulation outcomes). BNMs have found applications in the study of both healthy and pathological conditions. There are studies on resting-state, development and aging, but also on the effects of brain lesions, stroke, schizophrenia or dementia. Note that these works usually rely on results based on generic or average connectomes with some notable exceptions [6].

## 1.2 Large-scale patient specific brain models

BNMs based on the connectome can be extended to study the generation and propagation of epileptic seizures [8], and have been applied to the study of absence seizures [22], [23], temporal lobe epilepsy [4], and in a general framework for focal seizures, the Virtual Epileptic Patient (VEP) [5].

The first steps in the construction of a VEP model rely on the same elements described for a generic BNM, that is reconstruction of the connectome and choice of the neural mass model and coupling functions (figure 2). It is important that the reconstruction of structural connections is performed using state of the art methods to ensure that patient specific features are captured [5]. Under this requirement, there are reliable and reproducible differences in individual connectomes and the patient-specific connectome gives the best outcome for the VEP, where the biggest role is played by the topology of connections rather than by the weights [6]. Together with large-scale models for epilepsy based on the connectome [4], [5], there are other models that rely on functional connectivity (i.e. links represent temporal correlations among the activities of brain regions), as computed either from ictal or interictal recorded activity [7], [24], [25]. NMMs currently used in large-scale models for epilepsy are often phenomenological [4], [5], [7]. However, physiologically inspired (but still low-dimensional) NMMs have also been successfully used within this framework [25]. All these models contain a parameter for the excitability (or epileptogenicity), which describes how prone the brain region is to generate seizures. This is the key parameter that, together with the effect of the connectivity, marks the difference between healthy and epileptic brain tissue. The presence of brain lesions and clinical hypothesis on EZ and PZ, when available, can be included in the VEP by appropriate parameters modifications (figure 2). A hypothalamic hamartoma, for example, has been included by altering the local connectivity of the thalamus [5]. Clinical hypothesis on whether a brain region belongs to EZ, PZ or is healthy, can be set by tuning the value of the excitability parameter. The key feature of the model is the high degree of patient specificity.

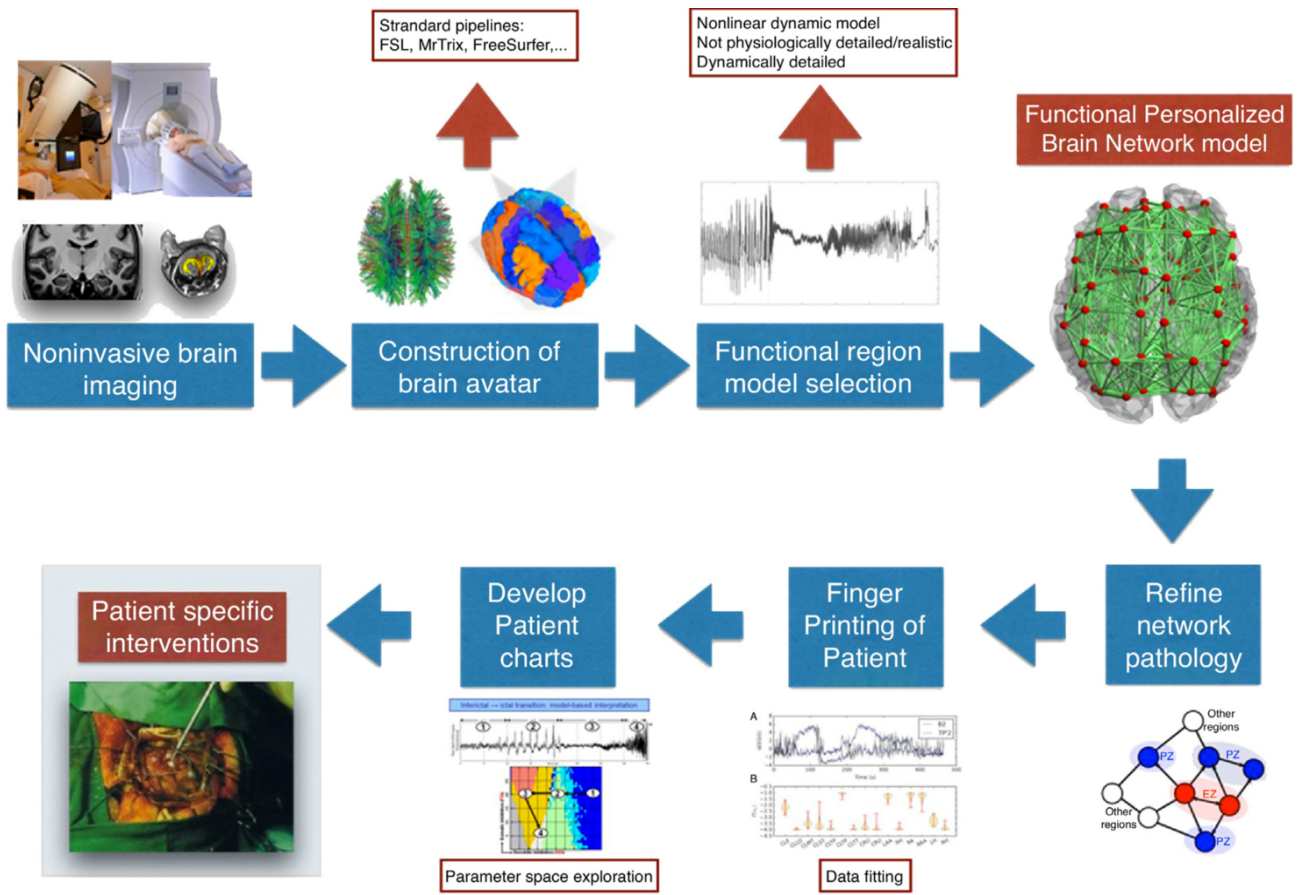

*Figure 2: The Virtual Epileptic Patient (VEP) Approach.* *A VEP model starts from the construction of the patient's brain avatar, based on non-invasive imaging. The second step is to choose the model for the dynamics of a single brain region. Currently, this choice does not depend on patient's specific seizure's characteristics. These steps bring to the functional characterization of the Brain Network Model (BNM,) after which the model can be further refined to consider differences in the excitability among regions, the presence of lesions or clinical hypothesis on the location of EZ and PZ. The model can be estimated through extensive data fitting to produce patients' charts containing information about the effects of specific modifications. These charts can be used to improve presurgical evaluation. Reprinted from Drug Discovery Today: Disease Models, 19, C. Bernard and V. Jirsa, Virtual Brain for neurological disease modeling, 5-10, Copyright (2016), with permission from Elsevier* [52].

## 2. Modeling seizures phenomenologically – NMMs

What model to use for the node dynamics? There is a large variety of models that have been proposed in the context of epilepsy and seizures [26]. They have been designed to investigate different types of epileptiform activity, including high frequency oscillations, spike-wave complexes, interictal spikes, fast oscillations at seizure onset and status epilepticus. As previously motivated, we focus here on phenomenological NMMs able to generate seizure activity. We will start by introducing some of the language of dynamical system theory useful to describe those models.

### 2.1 The language of dynamical system theory

In dynamical system theory the activity of a system, for example a neuron or a brain region, is described by a few variables, called *state* variables, such as the membrane potential, the mean firing rates of a set of neural populations or synaptic currents. The space spanned by the state variables is called state space, and the state of the system at a given time can be represented as a point in this space (figure 3A, left).

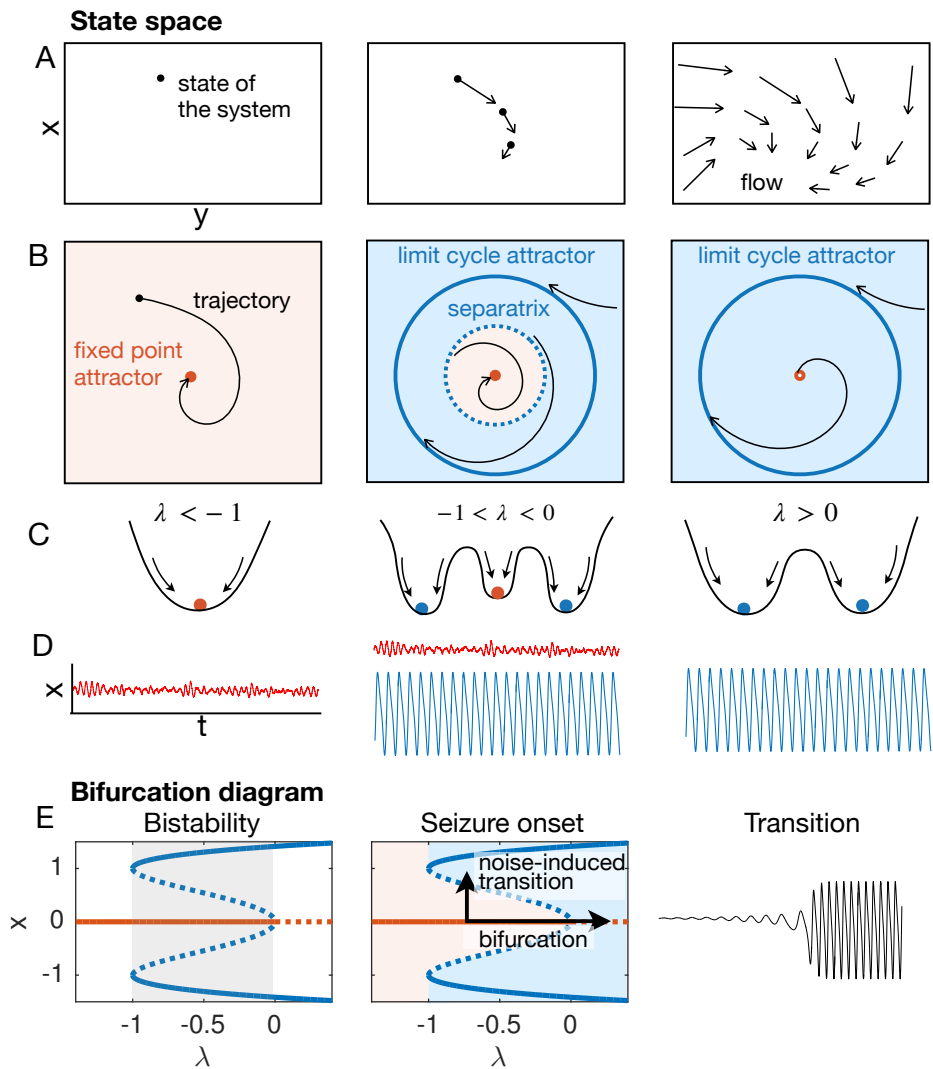

*Figure 3. Concepts from dynamical system theory.* If a system can be described using a few state variables (here two, (x,y)), its state can be described as a point in the (x,y) space, which is called state space (A, left). Differential equations describe how the system evolves in time starting form a given point (A, middle), this imposes a flow (A, right) in state space. Following the flow from a given point, we obtain a trajectory (B, left). Trajectories can be attracted to specific regions of state space, called attractors (B). An attractor can be a fixed point (B, left). For easier visualization we can imagine a valley, where gravity pushes the system down to the lowest point of the valley (C, left). If we slightly displace the system, this will move again down to the fixed point. An example of timeseries in the presence of small level of noise is shown in (D, left). Trajectories can converge to a closed orbit, a limit cycle attractor (C-D, right). This produces oscillatory activity. C, left should be visualized as a section of a Mexican hat, so that the two blue points are two points of the limit cycle. Attractors can also coexist (B-D middle). The basin of attraction of the limit cycle (shaded in blue) is separated from that of the fixed point (shaded in red) by a separatrix. A single system can display different attractors landscape when parameters (here lambda) are varied. When this occurs, the system has undergone a bifurcation. All the information in B can be represented in a compact form with a bifurcation diagram (E), where we plot one state variable versus the bifurcation parameter. The plot contains the attractors (full lines here) and the repellors, which are regions of state space that repel trajectories (dashed lines here). The coexistence of a limit cycle and a fixed point, called bistability (shaded in grey in E, left), is particularly relevant for models of seizure onset and offset. Starting from the interictal state (red fixed point), we can cross the separatrix changing the value of the state variable or that of the parameter (E, middle). An example of transition due to the latter mechanism is shown in E, right.

*Differential equations* encode the way in which state variables change over time, given a certain initial state of the system. They thus impose a vector field on state space (figure 3A, middle, right) and describe a movement in the state space, called a *trajectory* (figure 3B, left). After an initial transient, the system may settle in a specific behavior, such as steady or oscillatory, or even chaotic activities. In these cases, a portion of the state space, called an *attractor* (figure 3B), attracts trajectories passing from a specific region of the state space, the *basin of attraction*. An attractor can be a point in state space, as in the case of steady activity. This is called a stable *fixed point* (figure 3B, left). Oscillatory activity is exhibited by the system when the attractor is a closed trajectory, known as a stable *limit cycle* (figure 3B, right). *Strange attractors* are responsible for chaotic behaviors. Portions of the state space are called *repellors*, when they repel trajectories and comprise unstable fixed point and unstable limit cycles.

Differential equations can depend on some parameters, which are assumed to maintain a constant value while state variables evolve. In a neuron or NMM model they could encode, for example, the external applied current. When an attractor remains qualitatively the same for small changes of the parameters' values, it is said to be *structurally stable*. When, instead, these variations cause a change in the type and/or number of attractors, the system has undergone a *bifurcation* (figure 3B-D). The parameter(s) that one needs to modify to cause a bifurcation is called *bifurcation parameter(s)*. In the mentioned example of the neuron, this corresponds to the fact that applying an external current to a resting neuron does not trigger oscillations (the resting state is structurally stable) for a whole range of the current values. However, beyond the bifurcation point the system changes its state to an oscillatory one. We can make a plot of state variables versus bifurcation parameters containing all the possible attractors and repellors of a system. This is a *bifurcation diagram* (figure 3E, left, middle) and is a powerful compact description of the dynamic repertoire of the system and how this depends on the values of parameters and initial conditions.

The use of collective variables is particularly justified when modeling epileptic seizures. During seizures, in facts, the firing activity of billions of neurons becomes highly organized, which greatly reduces the degrees of freedom, hence the number of differential equations necessary to describe the observed activity (figure 4) [27]. Phenomenological models aim at identifying the essentials mechanisms able to produce the activity as observed, for example, in EEG recordings of seizures. They thus need to be able to generate sustained oscillatory activity, which require the use of at least two state variables so that a limit cycle attractor can exist. Since large-scale brain models for epilepsy aim at studying seizure onset and propagation patterns, most of these phenomenological models focus on the possible mechanisms to start a seizure, as predicted by dynamic system theory. Of note, multi-(spatial) scales recordings in epileptic patients point to different dynamical mechanisms for seizure onset acting at the population level as compared to the single neuron scale [28], further highlighting the importance of investigating and modeling emergent mesoscopic behaviors. Other features that have been modeled phenomenologically include seizure termination, spike and wave complexes, preictal spikes and status epilepticus. In the following subsections we will review models for the mentioned phenomena in more details.

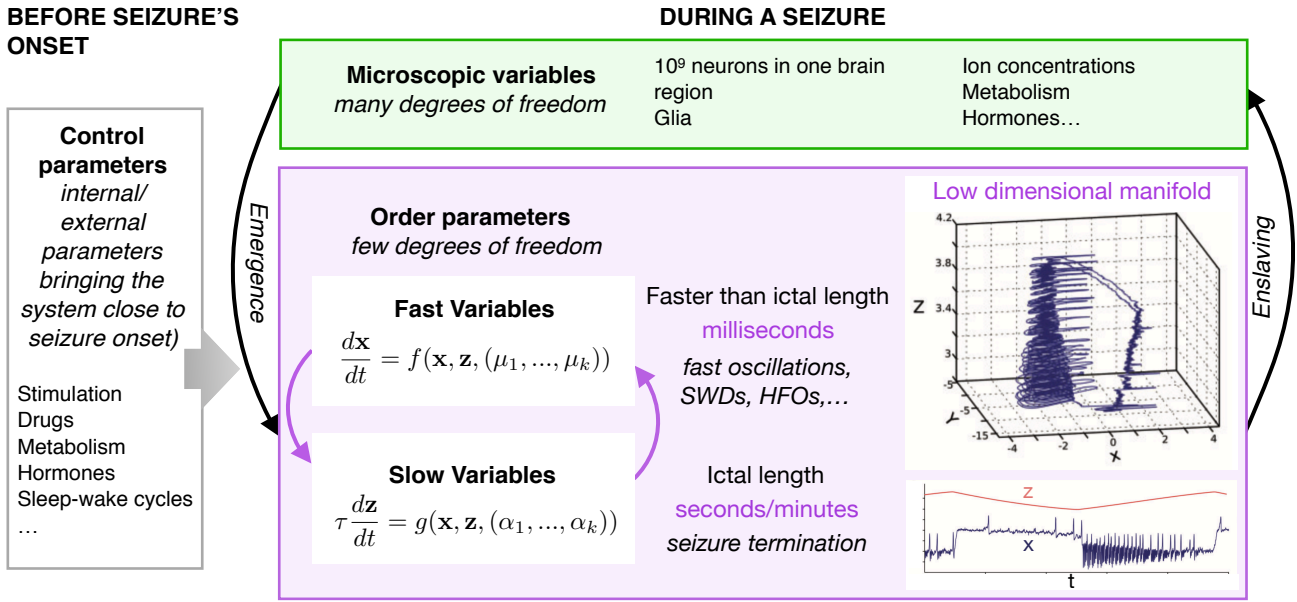

*Figure 4: Emergence of a low dimensional space with intrinsic time-scale separation during a seizure. Before a seizure, some (control) parameters of the brain change and bring the system close to seizure onset. During the seizure, the microscopic variables, which compose a brain region, organize so that the emergent global activity can be described by a few collective variables. We have thus a collapse of the many degrees of freedom of the system to a few emergent variables. These collective variables, on the other hand, act on the microscopic variables, 'enslaving' them. Some potential microscopic variables, which act as control parameters or on the slow time scale, are shown. The collective variables used to describe a seizure can be separated into at least two groups depending on the timescale at which they operate. Fast variables are responsible for the generation of a healthy 'rest' state and for an oscillatory 'seizure' state. Slow variables modulate the transition between these two states.*

## 2.2 Mechanisms for seizure onset and offset

Onset/offset mechanisms imply the transition between interictal to ictal states and vice versa. The interictal state is usually modeled as low amplitude fluctuations around a stable fixed point (or equilibrium), even though there exist models considering a non-quiescent interictal condition [29]. During the ictal state the system is usually in a stable limit cycle, but we will see an example where this does not occur. There are several possible onset dynamical mechanisms that have been identified in the literature [3], [9], [30], [31], some act at the level of a single node, while others rely on network effects (figure 5).

We will start with the single node. When the interictal (fixed point) and ictal (limit cycle) states coexists, as shown for example in (figure 3B-D, middle), we say we are in the presence of *bistability*. Looking at the bifurcation diagram of the bistable regime (figure 3E, left), we can identify at least two ways in which the system can leave the basin of attraction of the fixed point (shaded in red) and enter that of the limit cycle (shaded in blue) i.e. for a seizure to start (figure 3E, middle). In the first scenario, a change in the value of the state variable(s) (vertical axis), due to noise or to an internal/external stimulus, can cause the system to cross the separatrix. We will refer to this for simplicity as *noise-induced transition*. In the second scenario a change in the value of the parameter (horizontal axis) can bring the system in a condition where the fixed point is no longer stable, i.e. a *bifurcation* has occurred. Onset via a bifurcation can occur also without bistability. These are the most common mechanisms at the level of a single brain region, even though other theoretically possible ones have been proposed (e.g. attractors deformation [3], see figure 5).

Proposed mechanisms that depend on the effect of spatially extended networks are [30]: *excitability* and *intermittency*. In the first case, single nodes are not able to oscillate when isolated, but noise or internal/external stimuli can induce a single spike. When the nodes are coupled, the network can sustain oscillatory activity and a seizure is able to start and self-terminate. In the case of intermittency, heterogeneities in the network can create a new dynamical state at the level of the whole network that spontaneously produces brief seizures without the need of noise. Intermittency has only been observed in models for generalized seizures. More details are discussed in what follows.

### 2.2.1 Noise-induced transitions

Models in which transitions are due to movements in state space (a change in the value of the state variable(s)) usually rely on noise fluctuations to trigger seizures, but transitions can also be caused by an internal or external stimulus.

*Bistable models.* The simplest possible system with a bistability between a fixed point and a limit cycle shows the bifurcation diagram described in (figure 3E, left). For the readers familiar with bifurcations, this diagram arises in the normal form of the unfolding of the codimension-2 Bautin (or generalized Hopf) singularity. A model based on a modification of this normal form, called the $Z^6$ model [32], has been proposed as a phenomenological model for seizure generation that reproduced the relevant dynamics of a biophysically inspired NMM. In these family of models, seizures arise abruptly when noise fluctuations or inputs bring the system beyond the separatrix. The parameter $\lambda$ reflects the closeness to a bifurcation at which the fixed point loses stability, a subcritical Hopf bifurcation. The closest the value of $\lambda$ is to the bifurcation (i.e. to 0), the closest the separatrix is to the fixed point, so that $\lambda$ reflects the excitability of the system, in the sense of susceptibility to the generation of seizures. A greater excitability here translates in an average shorter escape time, that is the time employed by the system to make the transition due to noise fluctuations. Healthy brain regions can be modeled setting values of $\lambda$ further away from the bifurcation. If $\lambda$ is small enough (<-1) to be outside the bistability region, then the only possible attractor is the fixed point and seizures cannot be induced no matter how strong the input is. Seizure offset relies on the same noise-induced mechanism. However, for values of $\lambda$ that make a node more susceptible to generate seizures by making the separatrix closer to the fixed point, the distance from the separatrix to the limit cycle is high (see figure 3E, left). This implies large escape times for seizure offset, i.e. very long seizures, so that often seizure offset is not observed in simulations. Phenomenological models based on this simple bistable system have been used to build networks [2], [7], [24], [33], including a large-scale connectome-based model [4].

*Excitable models.* A system close to a bifurcation, i.e. an excitable system [34], can produce one or more oscillations if a stimulus is applied or as a consequence of noise fluctuations, before going back to the fixed point. In particular, a large amplitude oscillation can be produced by excitable systems close to what is called a SNIC bifurcation. When several nodes close to a SNIC bifurcation are coupled together, this transient oscillation triggered by noise can become sustained so that spontaneous seizures occur in the network. This was observed, for example, by using a physiologically inspired NMM as a node in a large-scale brain model [25]. The NMM includes the activity of four neural populations: one population of pyramidal neurons, one of excitatory interneurons and two populations of GABAergic interneurons. Depending on the values of its parameters it can also generate discharges through other mechanisms such as bistability, bifurcations and intermittency. Even though this model is not phenomenological, we include it here for two reasons: this physiologically inspired NMM is being used in large-scale patient specific models; excitability as an onset mechanism is amenable to phenomenological modeling.

### 2.2.2 Slow-variable dynamics and bifurcations

Seizures can be generated by changes in the parameters of the model that cause the disappearance of the fixed point in favor of a limit cycle. In this case we talk about bifurcations. Changes in the parameters can also cause the opposite transition, leading to seizure termination, or to transitions among different oscillatory states during a single seizure. Earlier physiologically inspired models exploiting this mechanism aimed at reproducing relevant features of the EEG as observed during seizure evolution, such as the appearance of spike and wave complexes [35]. This was done by manually changing the parameters to mimic the observed EEG patterns or by estimating them automatically from real data [36]. While this type of models, thanks to their explicit link to physiology, can offer powerful insights on the mechanisms underlying different electrographic patterns and the relations among them, they cannot spontaneously produce a seizure.

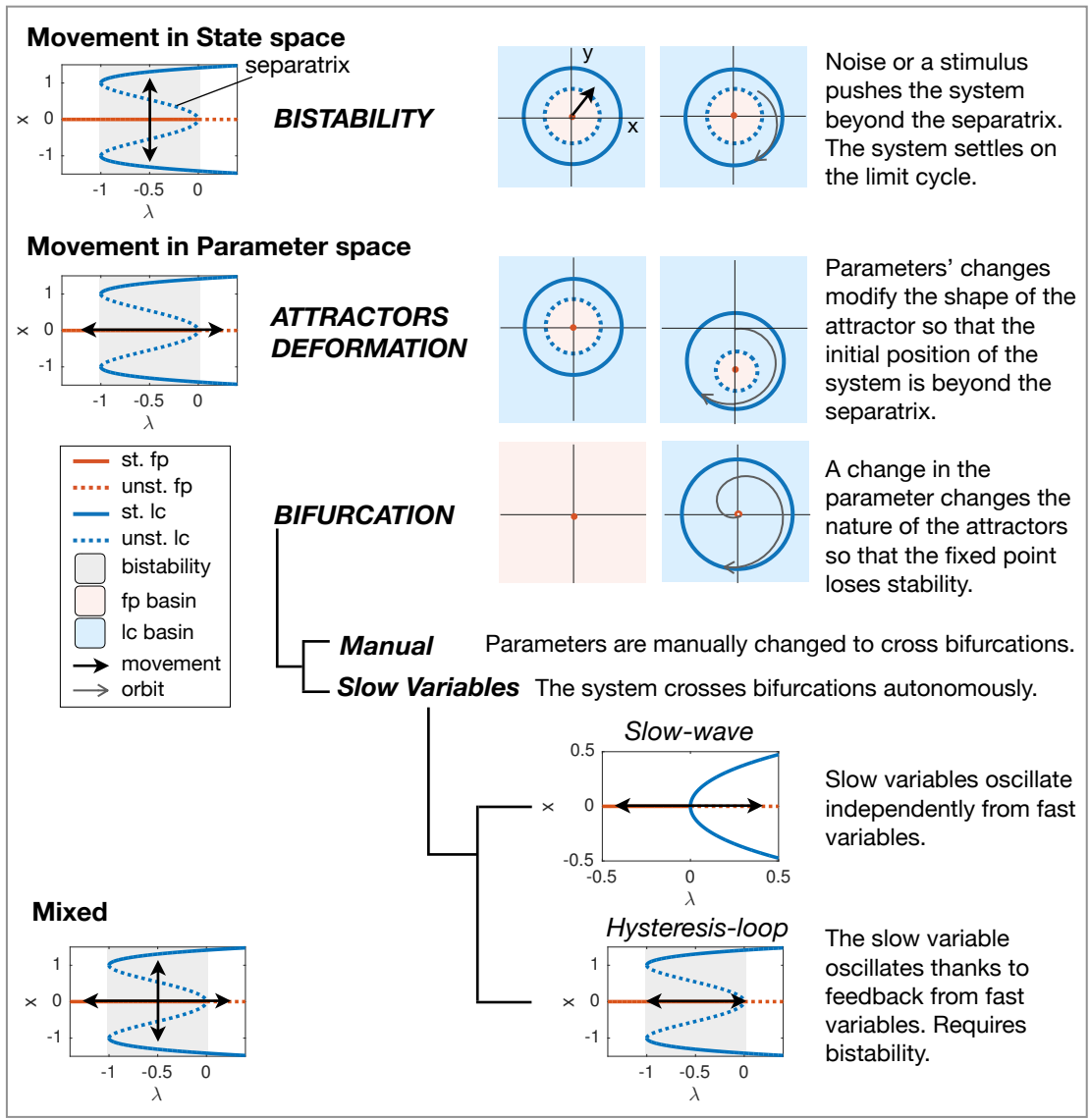

*Figure 5: Onset dynamical mechanisms. The upper panel shows mechanisms acting at the level of a single node, the lower panel those relying on the network effect. In the bifurcation diagrams, x is a variable and $\lambda$ the bifurcation parameter. For each mechanism, two plots of the state space (x, y) are portrayed: in the first we assume that the system is at rest in the stable fixed point (fp) and draw an arrow to show the movement in state space when it applies; in the second plot we show the orbit followed by the system to settle in the limit cycle (lc) or back on the fp. When movements in parameter space are performed, the two state space plots are for different values of the parameter.*

To create models able to autonomously generate/terminate seizures through bifurcations, we can consider that the bifurcation parameter(s) is itself a state variable(s) changing on a much slower timescale than that of the variables responsible for the limit cycle representing the ictal state. These latter variables will be called *fast variables* as opposed to the *slow variable* that brings the system across onset and offset bifurcations (figure 4). The timescale of the fast variables would be that of the oscillations within a seizure, while the slow timescale is that of the duration of a seizure. Slow variables can change in time following their own dynamics or their variation can depend on the activity of the fast variables (figure 5) [34]. While it is not clear which of these processes is more suitable to describe seizure onset, the distribution of ictal length's duration points to a negative feedback mechanism for seizure offset [27]. This implies that the ongoing seizure acts on the slow variables promoting a movement towards the offset bifurcation, as proposed in [9] within a phenomenological NMM for seizure activity, the Epileptor. Such a feedback mechanism requires the fast subsystem to exhibit bistability between the ictal and interictal states and can be implemented with the use of a single slow variable. The slow variable, called in [9] permittivity variable, is likely to encompass all the autoregulatory mechanisms triggered by the seizure, such as changes in extracellular ion concentrations or in variables related to energy metabolism, which eventually bring to seizure termination.

### 2.2.3 Mixed scenario with slow dynamics

The advantage of the permittivity variable approach is to provide a mechanism through which seizures can both start and terminate autonomously in the system. However, as mentioned before, while such a mechanism seems well justified for seizure offset, it is less clear whether it suits seizure onset. With regards to the latter, evidence from the analysis of the distribution of interictal lengths brings to mixed results [27]: seizures in vitro hippocampal preparation result to be highly periodic, and thus pointing to a permittivity variable approach for onset; seizures in rats models for absence seizures and in patients with focal or absence seizures bring to heterogeneous distributions, some of which are more consistent with noise-induced transitions. It is possible that deterministic (slow variables) and stochastic (noise-induced transitions) mechanisms coexist, with some being more prevalent in some patients than in others. Some types of epilepsies have a clear cyclicity, such as cathamenial epilepsy, in which the occurrence of seizures is linked to the menstrual cycle, or awakening seizures, in which seizures occur at a precise stage of the sleep-wake cycle. In general, though, such an evident cyclicity is lacking. Recently, however, it has been shown how the presence of circadian rhythms and clustering of seizures might be more ubiquitous than previously thought [37]. The analysis of long recordings (of the order of years) from patients with an implanted device for brain stimulation, showed that interictal epileptiform activity oscillates with both circadian and multi-diem rhythms, and that seizures preferentially occurred during specific phases of these cycles in a patient-specific fashion [37]. The authors suggested that such rhythms are co-modulated by different elements acting on several time scales, such as hormonal, genetic, environmental, sleep-wake cycle and behavior factors. These factors may act ultra-slowly (i.e. slower than ictal length) to periodically bring the system through the onset bifurcation or to alter brain excitability or circuitry, so that the separatrix gets closer to the fixed point and noise-induced transitions are facilitated. Interestingly, in models with a permittivity variable, the parameters can be set so that the onset can be noise-induced in a bistable regime and the offset due to slow movement in parameter space [9], [38], [39]. With the mathematical description proposed in [39], after seizure termination the system settles back to the fixed point with a value of the excitability much lower than it was at the beginning. In (figure 3E, left) this would be a value close to -1. These settings are thus compatible with the intriguing possibility that seizures serve as an emergency rescue mechanism that the system uses to restore a better working point.

### 2.2.4 Patient specific onset/offset bifurcations

While the slow variable mechanism can be applied to the bifurcation diagram described in the previous section (figure 3E, left), the Epileptor model differs in the type of bifurcations through which the fixed point/limit cycle loses stability (onset/offset bifurcation). There exist several bifurcations that can cause these transitions and they have different dynamical properties. Systems close to different bifurcations can react differently to an external stimulation or can have different synchronization properties which may affect seizure propagation (see [34], [39] and references therein). Given the clinical relevance of these properties, it is important to inform a seizure model with the best pair of onset/offset bifurcations able to describe data. This is made feasible by the fact that bifurcations may have specific signatures that can be identified in data, such as the behavior of the frequency or amplitude, or the presence of a jump in the signal baseline. In systems with two fast variables, there are only four possible onset bifurcations and four offset ones. This gives a taxonomy containing sixteen theoretically possible classes. Jirsa et al. [9] analyzed data from different species (zebrafish, mouse and human) and in vitro preparations and found that one specific class, having Saddle-Node (SN) onset and Saddle-Homoclinic (SH) offset, was predominant. This class is characterized by a jump in the baseline in DC (direct current) recordings and logarithmically decreasing frequency towards seizure offset. Interestingly, a jump in subdural EEG recordings has previously been found to be predictive of the seizure focus and patient outcomes [40]. The Epileptor model reproduces these dynamics. However, 20% of the patients analyzed showed different dynamics, which justified the creation of a model able to account for the other classes to improve the patient-specificity of the VEP [31]. This model predicted that seizures belonging to different classes could be produced by the same system. We have recently demonstrated, using a larger cohort of patients, that indeed several classes are necessary to describe the seizures analyzed and, furthermore, patients could have seizures of different type in time [39].

### 2.2.5 Excitability

In all the onset mechanisms described, it emerges the presence of a parameter which plays the role of excitability. This excitability parameter reflects the distance to the bifurcation that destabilizes the interictal state, and this holds both for mechanisms based on movement on state space, and on the crossing of the bifurcation point. While, from a biological point of view, seizures can be accompanied by both increased neuronal excitation and/or inhibition, in models both processes can contribute to an increased excitability, in the sense of closeness to instability point. This makes the brain region more susceptible to generating seizures. In large-scale brain models for epilepsy, fixing the values of this parameter allows to alter the local dynamics of the different brain regions involved in the network. In practice this can be done in different ways. For example, an increased excitability can reflect anatomical abnormalities of the brain region (such as atrophy [4]) or can reflect clinical hypothesis on the location of EZ and PZ [5].

### 2.2.6 When a seizure fails to terminate: status epilepticus

A failure in the mechanisms leading to seizure offset may result in a prolonged seizure (more than 5 minutes). This is a life-threatening condition known as status epilepticus. We are not aware of phenomenological models designed specifically to reproduce status epilepticus. Interestingly, though, dynamical states and mechanisms consistent with this condition have been found when investigating in greater detail some models for seizure onset and offset. Given their tractability, in facts, phenomenological models are amenable to a full exploration of the effects that changes in parameters' values have on the landscape of attractors. This may bring to uncovering novel behaviors that were not intentionally included in the model. The more canonical the model, the likelier such 'unintentional' behaviors are to survive the empirical test and their presence can foster our understanding of the full potential of the system we are studying.

With regards to status epilepticus, two hypotheses have emerged: (i) status epilepticus as an additional attractor coexisting with the ictal one or (ii) as a consequence of a failure in the slow variable mechanisms. The first hypothesis arises from a detailed exploration of the bifurcation diagram of the Epileptor model, which led to the discovery of other attractors that are reminiscent of clinically relevant behaviors, including status epilepticus [41]. The authors provided first experimental support for the existence of these attractors by performing experiments, which trace out the paths in parameter space through experimental manipulations, in particular forcing the hippocampus (rodent, in vivo) through a sequence of different behaviors as predicted from the bifurcation diagram. Another phenomenological model, instead, found a behavior resembling status epilepticus when, in a specific region of the bifurcation diagram, the slow variable's attempts of bringing the system towards seizure termination were overridden by excessive noise [39]. This is in line with a dynamical hypothesis for status epilepticus previously formulated in the context of a physiologically inspired NMM [28], in which the system repeatedly approaches the critical transition without crossing it and retreats to the ictal attractor. In vivo data (EEG and intracranial EEG) in the latter study showed changes in the frequency, mean power and mean autocorrelation similar to those predicted by the model for this dynamical mechanism.

## 2.3 Seizure evolution: a dictionary of EEG patterns

### 2.3.1 Navigating the parameter space

The presence of multiple timescales within the variables responsible for the oscillatory activities observed during a seizure is at the heart of the generation of a variety of rhythms in physiologically inspired NMMs [42]. In particular, at least three variables with two timescales are necessary to produce complex activity such as spike and wave complexes or polyspike waves.

Wang and colleagues [42] proposed a minimal model of three generic neural processes acting on two timescales to reproduce different prototypical EEG patterns, as observed during seizure evolution, when the model's parameters vary. The model exhibits interictal activity as fixed point dynamics. Bifurcations can cause the appearance of fast sinusoidal oscillations and spike trains, which are represented in the model in terms of simple limit cycles in the fast subsystem. Interaction with the slower variable produces slow waves, with different numbers of spikes riding on it. Note that this slower variable is still much faster than that leading to seizure offset. The authors suggest that it could be related to processes such as the regulation of extracellular potassium, glial processes, or the effect of subcortical input. Manually changing the parameters of the model allows to navigate the map of possible attractors of the system (i.e. the bifurcation diagram), mimicking the action of slow variables and reproducing the realistic successions of EEG patterns observed during a seizure. The model offers a variety of behaviors that can be used as a dynamic 'dictionary' to help identifying relevant activities, and relationships among them, in more complex physiologically inspired NMMs or in data [43].

### 2.3.2 A single mechanism for spike and waves discharges and interictal spikes

The addition of an intermediate timescale between fast and slow ones is responsible for the generation of spike and wave complexes also in another phenomenological model, the Epileptor. Here the same mechanism is responsible also for the generation of preictal spikes. When the fast variables approach seizure onset, the intermediate system approaches a SNIC bifurcation: the average of the fast variables plays the role of bifurcation parameter for the intermediate system. Interestingly, this implies that the closer the fast variables are to seizure onset, the more likely are the intermediate variables to show isolated spikes due to noise fluctuations, consistent with the experimental observation of preictal spikes. In contrast to the fast subsystem of the model, this mechanism is ad hoc and needs to be rooted in empirical data [9]. It accounts for the fact that the slow variable is stochastic and allows for phenomena such as that the increase of the number of interictal spikes is not necessarily followed by a seizure. During the seizure, the intermediate variables produce a sharp oscillatory activity that interact with the fast oscillations of the fast variables to create spikes and wave complexes, in a similar fashion to Wang et al. [42], but involving different bifurcations. These complexes are particularly evident as the seizure progress.

This modeling work allows us to discuss two further points. The first is about the interpretation of the groups of variables composing a phenomenological model, that in biophysically inspired NMMs are typically related to different neural populations, each composed of similar neurons. Experimental results from whole cell hippocampi recordings aimed at identifying the biophysical correlates of fast and intermediate Epileptor variables, showed that, while the activity of the intermediate variables was more representative of GABAergic cells, pyramidal cells contributed as well. The interpretation of these sets of collective variables may be more complex than linking each of them to either an excitatory or inhibitory cells population and phenomenological NMMs can encode the activity of non-homogeneous neural populations .

Secondly, this provides an example of how specific dynamical features can sometimes be modeled in isolation. While intermediate subsystem contributes to our understanding of seizure related phenomena and makes the simulated timeseries more realistic, it was shown by the authors that this subsystem does not contribute to seizure onset and offset mechanisms. Successive work could thus be simplified by omitting the modeling of interictal spikes and spike and wave complexes, without losing any understanding on the mechanisms underlying seizure initiation and termination.

# 3. Applications and future directions

## 3.1 In silico experiments to improve treatment

### 3.1.1 Virtual surgeries and stimulations

Large-scale models for epilepsy can find several important applications to improve treatment (figure 2). A first promising use is a better delineation of the EZ and PZ. This can be achieved by testing whether the clinical hypothesis brings to simulated functional data that match the empirical ones, for example in terms of seizure onset and propagation pattern. Parameters in models can be changed iteratively to reduce the mismatch [5]. When several clinical hypotheses have been formulated, the large-scale patient model can help to choose among them. In addition, it can be used to estimate EZ and PZ through Bayesian approaches, without prior hypothesis [5]. Of note, all these approaches to the delineation of EZ and PZ can be applied starting from non-invasive recordings and can be used to improve the placement of SEEG electrodes.

Large-scale models for epilepsy also provide a platform to perform in silico surgeries. Once the EZ has been delineated, one could test if resection would stop seizure propagation. Results could help improving the resection strategy [4], [5], [25] or could help understanding when the outcome of surgery will be unsatisfactory [7]. Even though these models have only been evaluated retrospectively, a mismatch between the resection predicted by the model and real resection correlates with poor surgical outcome [5], [25]. These models can also help proposing innovative surgical approaches, which fully exploit the network nature of this disease, such as micro lesions or multiple lesions at different locations that could make use of recent stereotactic-guided laser technology [5]. This would provide a less invasive approach for network control [44]. Finally, a minimal resection is not necessarily the best strategy, as it may involve eloquent brain areas leading to post-surgery neurological complications. An exciting possibility is to use large-scale simulations to identify, among the possible efficient resections, those that are safest in terms of normal brain function [45].

In silico experiments could include the design of patient specific stimulation protocols to prevent or abort seizures (see [46] for a review). Stimulation could be designed, for example, to force the system out of the basin of attraction of the limit cycle (the seizure) in a bistable regime, where the timing and amplitude of the stimulus necessary to abort the seizure would depend on the shape of the basin of attraction of the healthy state. Stimulations can also modify the bifurcation structure of the system or modulate its excitability to prevent seizure onset. Combining the local node dynamics with large-scale connectivity could bring, again, to a less focal and more distributed intervention approach.

As reviewed in the previous section, the dynamics of a single brain region in current large-scale patient specific models can rely on different dynamical mechanisms: (i) noise-induced transitions in a bistable regime created by a subcritical Hopf bifurcation in phenomenological models of the Bautin family (normal form of the Bautin singularity, Z6 model…); (ii) slow variable induced bifurcations in phenomenological models (Epileptor); or (iii) excitability close to a SNIC bifurcation in a physiologically inspired model, in which the mechanism for seizure onset relies on the interplay between the node's excitability and the whole network activity (an isolated node cannot sustain oscillations). The way in which different brain regions influence each other varies in the literature, including linear and diffusive couplings, with or without time delays [4], [7], [24] and couplings across timescales [5], [6], [47]. The large-scale models also differ in the way the connectivity among brain regions is assessed (from structural or functional data, and in the latter case using ictal or interictal recordings), on the coverage (some brain regions or the whole brain) and in how the best treatment strategy is evaluated. Patient-specific resections can be suggested based entirely on brain network simulations or mathematical analysis. The most ictogenic nodes can be identified, for example, as those with shorter escape times (i.e. those generating a seizure more easily) [7], or those that, when removed from the network, decrease the occurrence of seizures [25]. Linear stability analysis applied to a combination of structural and dynamical information can be used to estimate the propagation zone and suggest minimal ablations in the connectivity to stop seizure recruiting [44]. In addition to connectivity and dynamics, one can also take advantage of the availability of functional data (EEG, SEEG, MEG…) to apply data fitting techniques for the estimation of the excitability of each network node through model inversion [5].

Despite their differences, all these studies have brought positive results, while showing that network measures alone are not predictive and that dynamical models are essentials [4], [44]. Large-scale patient specific model approach, thus, holds promises to help clinicians in their decisions about electrodes placement and surgery interventions. Currently, positive results do not seem to strictly depend on the specific choices in terms of modeled dynamics and connections. This may be due to the use of different metrics to assess the predictive power of the approach. Future work may address under which circumstances the model's choice will affect the propagation pattern of seizures and what are the best domains of applications of the different models; open to the possibility that patient specificity may express itself also with the need of choosing one model over another.

For example, we have recently shown in a large cohort of patients that different bifurcations are needed to explain the variability at onset and offset among patients, and even in the seizures of a single patient. There is evidence in the literature that different bifurcations may greatly affect the synchronization properties of the nodes. A more systematic investigations of these properties, their effects on seizure recruitment and their dependence on the type of coupling, would help to understand how much dynamical realism is necessary to include in the phenomenological model for a specific study. As for physiologically inspired models, for which one has to decide, depending on the specific question addressed and data available, what is the degree of biological realism needed (single neurons? dendritic branches? ion channels properties? ...), similarly, for phenomenological models one has to choose the level of dynamical realism. Further studies may address for instance when bistability is enough, when the specific bifurcations giving origin to the bistability or being directly responsible for seizure onset and offset matter, whether different mechanisms (noise-induced transitions, bifurcations with slow dynamics, excitability, intermittency…) coexist, what is the interplay among them and when one should choose one over the other. An explicit example is the coexistence between the slow variable mechanism, which can increase the excitability of a brain region by bringing it closer to an onset bifurcation, and noise-induced transitions that are facilitated with the increased excitability. This interplay has been suggested by several authors and, under some assumptions (e.g. timescales separation), can be used to justify the use of noise-induced transitions at seizure onset, while the slow dynamics is approximated by giving different excitability values to the nodes of the system.

Finally, beyond applications that can be directly used to improve patient-specific clinical care, these large-scale simulations can foster our understanding of the link between structure and function in epilepsy and seizure propagation [5][5] or of the conditions that facilitate seizure's genesis and propagation. A complex system approach can help to advance counterintuitive hypotheses. For example, a general teaching is that, for a seizure to occur, we need: (i) an altered excitability of the single units involved (neurons, brain regions...) and (ii) an altered network connectivity which promotes propagation and/or synchronization among units. However, it has been shown, on the basis of computational studies, that a pathological connectivity can even provoke seizures in networks composed only by 'healthy' units [2]. It is then possible that the two requirements, of pathological units and pathological connectivity, need not to be necessarily satisfied independently, but that it is the interplay between them that can bring to a pathological condition.

### 3.1.2 New strategies out of seizures

Virtual surgeries and stimulations are the most straightforward, although highly non trivial, applications of meso- and large-scale phenomenological models in epilepsy. Another intriguing possibility is to use the bifurcation diagrams of mesoscopic models (both physiologically inspired or phenomenological) as a map to guide the system out of 'dangerous' regions. A dangerous region would be any portion of the diagram in which the landscape of attractors facilitates seizures, as, for example, when the system gets very close to an onset bifurcation or to status epilepticus. A safer region would be far from such bifurcations, ideally outside the bistability region so that the interictal state is the only existing attractor.

However, this require acting on the variables and parameters of the bifurcation diagram, which poses at least two challenges: identifying what these variables and parameters are in the real system and finding tools to manipulate them. This task may be particularly difficult when dealing with phenomenological models, since the link between the model's variables and the real biological substrate is not evident. At the same time, this same type of models is particularly powerful in providing tractable bifurcation diagrams.

Strategies to find correlates also depend on how the phenomenological model was created. As shown in figure 1, one possible way to build such a model is to start from a biophysical inspired one, identify the relevant dynamics one wants to study (as represented, for example, in the bifurcation diagram) and create a simpler model (possibly the simplest) able to reproduce it. A classic example in neural modeling is the FitzHugh-Nagumo model, created to isolate the essential mechanisms underlying the dynamics of action potentials in the Hodgkin-Huxley model and to have a more tractable description. In the context of mesoscopic seizure models, we mentioned the Z6 model as an easier version of a physiologically inspired one. When a formal mathematical reduction is available, this provides a link between variables in the physiologically inspired and phenomenological models, where variables in the latter are typically a function of one or more physiological ones. More in general, a link can be suggested also by comparing variables and parameters appearing in the bifurcation diagrams.

Another class of phenomenological models (figure 1), instead, aims at reproducing directly specific features of the activity observed in experimental data. The Epileptor model is an example of this approach. In these cases, one could make hypotheses on the physiological correlates of the variables, based, for example, on the timescale at which the process acts, and try to test them experimentally. Another possibility is to compare the model with physiologically inspired ones that exhibit a similar bifurcation diagram a posteriori. However, it is important to bear in mind that a phenomenological model could have a variety of physiological implementations. In the case of seizures, Jirsa and colleagues have shown how the specific dynamics they modeled in the Epileptor was conserved at different scales, from small networks of neurons to brain regions, and across species, from flies and zebra fish to mice and humans. This suggests that there are several substrates that can support the modeled activity. This could hold also for patients, given the considerable diversity of causes and conditions that can bring to epilepsy.

## 3.2 Dynamics as a criterion to classify seizures

Keeping with the idea that there is not a single dynamical mechanism that can fully explain all seizures observed in patients, details about this mechanism could contribute to the classification of electrographic seizures. And in this, phenomenological models would help to isolate the essential dynamics. Current seizure classifications, in facts, are based on a description of the empirical data: clinical manifestations (e.g. partial vs. generalized) together with visual descriptions of EEG signal and identification of the regions of the brain involved. In one of its position papers, the ILAE explains that, while several ways of classifying seizures could be possible, due to the lack of fundamental knowledge, the current classification is practical and based on previous classifications. Among the criteria that could be used to classify seizures are mentioned: pathophysiology, anatomy, networks involved, and practical criteria such as response to antiepileptic drugs, EEG patterns, level of related cognitive and physical impairment. An additional criterion has been recently proposed [9], [48], that is distinguishing electrographic seizures based on the dynamical mechanisms leading to onset and, possibly, offset. One of the advantages of this approach is that it highlights differences in the underlying mechanisms of seizure generation, evolution and termination, which could contribute to our understanding of this phenomenon. Another advantage is that models for the different dynamics could be used to make useful predictions with applications in the clinical setting [49], as described in previous sections. There are currently two main proposals, that are highly complementary. One focuses on spatial heterogeneities in the excitability within a single brain region [48], which can produce either low amplitude fast activity (LAF) or high amplitude slow activity (HAS) at seizure onset through different dynamical mechanisms (involving local and global bistability, monostability, stimulus induced transitions and bifurcations). This modeling approach is powerful in capturing the contribution of the local spatiotemporal dynamics within a brain region and the authors show how the latter onset type is linked to worse surgical outcomes when a portion of the brain region is removed, in accordance with clinical results [50]. The second proposal focuses on the possible bifurcation scenarios at the level of a NMM [9], where different seizure classes have different onset/offset bifurcations pairs. Both approaches bear the potentiality of bringing to useful predictions for the clinics, and future work could be done to propose a multi-scale classification. In general, we could say that a taxonomy of seizures based on dynamics could include the specification of any type of dynamical process leading to seizure onset, evolution and offset. This would powerfully complement current operational classifications based on clinical information.

This could be applied also to different brain regions in a single patient. Is the difference between those in EZ, PZ or healthy ones just a matter of altered excitability as it is usually assumed, or are there deeper differences in terms of dynamical mechanisms? The latter hypothesis would translate in the need of different models for the various regions' types.

## 4. Conclusions

Epilepsy is a complex network and dynamic disease, involving several interacting spatial and time scales. Efforts to understand the mechanisms underlying the generation, evolution and termination of epileptic seizures and to improve the life of epileptic patients benefit from the use of diverse complementary approaches. In this chapter we have reviewed phenomenological models of seizure-like activity. We have highlighted the advantages of this modeling approach, the types of questions it can be used to address and how it has been applied in the investigation of epileptic seizures. Efforts in the field have focused mainly on identifying dynamical mechanisms responsible for seizure onset and offset. Systems able to autonomously generate, and possibly terminate, a seizure, in facts, are at the heart of large-scale patient specific brain models, which have the potential of improving clinical care for drug resistant epileptic patients. These virtual epileptic patient models can be used not only as an additional tool to delineate the best treatment strategy within the realm of current clinical practice but can suggest innovative scenarios. Examples are distributed and minimally invasive ablations or stimulations that fully exploit the network and dynamical properties of the system, or even modulation of the slow variables and parameters to force the system in safer regions of the bifurcation diagram. Applications are currently validated at best retrospectively on data from patients who have already undergone surgery, to help explaining surgery's outcomes and propose better interventions. The target is a set of tools that could routinely help clinicians in making the best choice for each patient.

Beyond these applications of more immediate translational value, phenomenological models can help to foster our understanding of the mechanisms underlying epileptic seizures. The characterization of their essential dynamics could promote a taxonomy of seizures which bears predictive values for issues relevant for treatment.

All the phenomenological models considered in this review have the same spatial scale -a brain region-. At the level of a single brain region there are three main ingredients that characterize the dynamics of the model and allow to understand the relationship among different models.

The first is the bifurcation structure of the model, which determines the range of possible behaviors of the system. This is a key ingredient even when bifurcations do not occur in the model, since it nevertheless determines the excitability properties of the system.

The second ingredient is the role of what we here generically called 'noise', which can include both internal or external fluctuations and stimulations. Noise can cause transitions among co-existing attractors, bringing to seizure onset or offset, also in the presence of slow variable mechanisms; it can cause preictal spiking when the system gets closer to an onset bifurcation; combined with the network effect, it can sustain oscillations in excitable systems which are not able to oscillate in isolation; and excessive noise might prevent seizure termination.

The third ingredient is the existence of several timescales having distinct roles in seizure dynamics (figure 4). A fast timescale is responsible for the oscillatory activity observed during a seizure. It groups phenomena acting on different timescales themselves, such as fast oscillations or spike and wave complexes. Despite this heterogeneity, all of them are much faster than the typical seizure duration. This fast timescale is typically linked to the electrical activity of neural populations, which produces oscillatory mesoscopic activity, but slower processes such as regulation of extracellular potassium, glial processes, or the effect of subcortical input contribute to shaping the different waveforms. These oscillations trigger slow microscopic processes that cooperate to terminate the seizure. The slow timescale is thus comparable to the ictal length. Interestingly, these processes can also be described by a collective variable, the permittivity, which emerges from the balance (or imbalance) of pro- and anti-seizures mechanisms pushing the fast system towards seizure offset. We thus have emergent collective variables exhibiting intrinsic timescales separation, with different scales influencing each other. This thinking is novel from the theoretical point of view. We can thus say that during an epileptic seizure the activities of billion of neurons, but also that of extracellular ion concentrations, glia, variables related to metabolism and so on, become organized so that a few collective variables emerge that capture the system dynamics. This implies a collapse of the degrees of freedom of the system, and the repertoire of possible activities can be summarized using a low dimensional manifold and the flow on it as induced by the landscape of attractors/repellors. Processes promoting the collapse on this 'seizure manifold', typically acting on an ultra-slow timescale such as neuromodulators, hormones, variables linked to the sleep-wake cycle and so on, are those responsible for bringing the fast system close to the onset bifurcation. These processes would also be responsible for the expression of one seizure class over another, as described in previous sections.

It has been proposed that the ultra-slow and slow dynamics identified in the context of seizures' modeling exist and play an equally important role in the healthy brain [51]. This hypothesis comes from the physiological evidence that any brain can exhibit seizures under the right conditions, so that they might be part of the dynamical repertoire of any brain together with other brain rhythms. The above description about the emergence of two-tiered order parameters (collective variables) is, in facts, reminiscent of the Structured Flows on Manifold framework proposed as a formal description of behavioral and brain organization, characterized by the existence of at least two coevolving timescales in the order parameters linked to the emergence of behavior [51]. In particular, fast variables express the execution of cognitive and action tasks, while slow variables modulate the creation and annihilation of the specific low dimensional space, and attractor landscape therein, in which a task can occur, in a similar fashion to the mechanism causing the collapse on the low dimensional space on which seizure activity takes place.

# References

1. Fisher RS, Boas W van E, Blume W, et al. Epileptic seizures and epilepsy: definitions proposed by the International League Against Epilepsy (ILAE) and the International Bureau for Epilepsy (IBE). Epilepsia. 2005; 46(4):470–2.
2. Terry JR, Benjamin O, Richardson MP. Seizure generation: the role of nodes and networks. Epilepsia. 2012; 53(9).
3. Da Silva FL, Blanes W, Kalitzin SN, et al. Epilepsies as dynamical diseases of brain systems: basic models of the transition between normal and epileptic activity. Epilepsia. 2003; 44(s12):72–83.
4. Hutchings F, Han CE, Keller SS, et al. Predicting surgery targets in temporal lobe epilepsy through structural connectome based simulations. PLoS Comput Biol. 2015; 11(12):e1004642.
5. Jirsa VK, Proix T, Perdikis D, et al. The virtual epileptic patient: individualized whole-brain models of epilepsy spread. Neuroimage. 2017; 145:377–88.
6. Proix T, Bartolomei F, Guye M, et al. Individual brain structure and modelling predict seizure propagation. Brain. 2017; 140(3):641–54.
7. Sinha N, Dauwels J, Kaiser M, et al. Predicting neurosurgical outcomes in focal epilepsy patients using computational modelling. Brain. 2016; 140(2):319–32.
8. Taylor PN, Kaiser M, Dauwels J. Structural connectivity based whole brain modelling in epilepsy. J Neurosci Methods. 2014; 236:51–7.
9. Jirsa VK, Stacey WC, Quilichini PP, et al. On the nature of seizure dynamics. Brain. 2014;

137(8):2210–30.
10. Südhof TC. Molecular Neuroscience in the 21st Century: A Personal Perspective. Neuron. 2017.
11. Eliasmith C, Trujillo O. The use and abuse of large-scale brain models. Curr Opin Neurobiol. 2014; 25:1–6.
12. Jirsa VK, McIntosh AR. Handbook of brain connectivity. Vol. 1. Springer; 2007.
13. Deco G, Jirsa V, McIntosh AR, et al. Key role of coupling, delay, and noise in resting brain fluctuations. Proc Natl Acad Sci. 2009; 106(25):10302–7.
14. Breakspear M. Dynamic models of large-scale brain activity. Nat Neurosci. 2017; .
15. Sanz-Leon P, Knock SA, Spiegler A, et al. Mathematical framework for large-scale brain network modeling in The Virtual Brain. Neuroimage. 2015; 111:385–430.
16. Kalitzin S, Koppert M, Petkov G, et al. Multiple oscillatory states in models of collective neuronal dynamics. Int J Neural Syst. 2014; 24(06):1450020.
17. Naze S, Bernard C, Jirsa V. Computational modeling of seizure dynamics using coupled neuronal networks: factors shaping epileptiform activity. PLoS Comput Biol. 2015; 11(5):e1004209.
18. Eissa TL, Dijkstra K, Brune C, et al. Cross-scale effects of neural interactions during human neocortical seizure activity. Proc Natl Acad Sci. 2017; 114(40):10761–6.
19. Proix T, Jirsa VK, Bartolomei F, et al. Predicting the spatiotemporal diversity of seizure propagation and termination in human focal epilepsy. Nat Commun. 2018; 9(1):1088.
20. Iturria-Medina Y. Anatomical brain networks on the prediction of abnormal brain states. Brain Connect. 2013; 3(1):1–21.
21. Proix T, Spiegler A, Schirner M, et al. How do parcellation size and short-range connectivity affect dynamics in large-scale brain network models? Neuroimage. 2016; 142:135–49.
22. Taylor PN, Goodfellow M, Wang Y, et al. Towards a large-scale model of patient-specific epileptic spike-wave discharges. Biol Cybern. 2013; 107(1):83–94.
23. Yan B, Li P. The emergence of abnormal hypersynchronization in the anatomical structural network of human brain. Neuroimage. 2013; 65:34–51.
24. Benjamin O, Fitzgerald THB, Ashwin P, et al. A phenomenological model of seizure initiation suggests network structure may explain seizure frequency in idiopathic generalised epilepsy. J Math Neurosci. 2012; 2(1):1.
25. Goodfellow M, Rummel C, Abela E, et al. Estimation of brain network ictogenicity predicts outcome from epilepsy surgery. Sci Rep. 2016; 6:29215.
26. Wendling F, Benquet P, Bartolomei F, et al. Computational models of epileptiform activity. J Neurosci Methods. 2016; 260:233–51.
27. Suffczynski P, Da Silva FHL, Parra J, et al. Dynamics of epileptic phenomena determined from statistics of ictal transitions. IEEE Trans Biomed Eng. 2006; 53(3):524–32.
28. Kramer MA, Truccolo W, Eden UT, et al. Human seizures self-terminate across spatial scales via a critical transition. Proc Natl Acad Sci. 2012; 109(51):21116–21.
29. Goodfellow M, Schindler K, Baier G. Intermittent spike--wave dynamics in a heterogeneous, spatially extended neural mass model. Neuroimage. 2011; 55(3):920–32.
30. Baier G, Goodfellow M, Taylor PN, et al. The importance of modeling epileptic seizure dynamics as spatio-temporal patterns. Front Physiol. 2012; 3:281.
31. Saggio ML, Spiegler A, Bernard C, et al. Fast--Slow Bursters in the Unfolding of a High Codimension Singularity and the Ultra-slow Transitions of Classes. J Math Neurosci. 2017; 7(1):7.
32. Kalitzin S, Koppert M, Petkov G, et al. Computational model prospective on the observation of proictal states in epileptic neuronal systems. Epilepsy Behav. 2011; 22:S102--S109.
33. Goodfellow M, Taylor PN, Wang Y, et al. Modelling the role of tissue heterogeneity in epileptic rhythms. Eur J Neurosci. 2012; 36(2):2178–87.
34. Izhikevich EM. Neural excitability, spiking and bursting. Int J Bifurc Chaos. 2000; 10(06):1171–266.
35. Wendling F, Bartolomei F, Bellanger JJ, et al. Epileptic fast activity can be explained by a model of impaired GABAergic dendritic inhibition. Eur J Neurosci. 2002; 15(9):1499–508.
36. Wendling F, Hernandez A, Bellanger J-J, et al. Interictal to ictal transition in human temporal lobe epilepsy: insights from a computational model of intracerebral EEG. J Clin Neurophysiol. 2005; 22(5):343.
37. Baud MO, Kleen JK, Mirro EA, et al. Multi-day rhythms modulate seizure risk in epilepsy. Nat Commun. 2018; 9(1):88.
38. Meisel C, Kuehn C. Scaling effects and spatio-temporal multilevel dynamics in epileptic seizures.

PLoS One. 2012; 7(2):e30371.
39. Saggio ML, Crisp D, Scott J, et al. A taxonomy of seizure dynamotypes. *bioRxiv*. 2020;
40. Ikeda A, Taki W, Kunieda T, et al. Focal ictal direct current shifts in humanepilepsy as studied by subdural and scalp recording. Brain. 1999; 122(5):827–38.
41. El Houssaini K, Ivanov AI, Bernard C, et al. Seizures, refractory status epilepticus, and depolarization block as endogenous brain activities. Phys Rev E - Stat Nonlinear, Soft Matter Phys. 2015; .
42. Wang Y, Goodfellow M, Taylor PN, et al. Phase space approach for modeling of epileptic dynamics. Phys Rev E. 2012; 85(6):61918.
43. Touboul J, Wendling F, Chauvel P, et al. Neural mass activity, bifurcations, and epilepsy. Neural Comput. 2011; 23(12):3232–86.
44. Olmi S, Petkoski S, Guye M, et al. Controlling seizure propagation in large-scale brain networks. PLoS Comput Biol. 2019; .
45. An S, Bartolomei F, Guye M, et al. Optimization of surgical intervention outside the epileptogenic zone in the Virtual Epileptic Patient (VEP). PLoS Comput Biol. 2019; 15(6):e1007051.
46. Wang Y, Hutchings F, Kaiser M. Computational modeling of neurostimulation in brain diseases. In: Progress in Brain Research. 2015.
47. Proix T, Bartolomei F, Chauvel P, et al. Permittivity coupling across brain regions determines seizure recruitment in partial epilepsy. J Neurosci. 2014; 34(45):15009–21.
48. Wang Y, Goodfellow M, Taylor PN, et al. Dynamic mechanisms of neocortical focal seizure onset. PLoS Comput Biol. 2014; 10(8):e1003787.
49. Wang Y, Trevelyan AJ, Valentin A, et al. Mechanisms underlying different onset patterns of focal seizures. PLoS Comput Biol. 2017; 13(5):e1005475.
50. Doležalová I, Brázdil M, Hermanová M, et al. Intracranial EEG seizure onset patterns in unilateral temporal lobe epilepsy and their relationship to other variables. Clin Neurophysiol. 2013; 124(6):1079–88.
51. McIntosh R, Jirsa V. THE HIDDEN REPERTOIRE OF BRAIN DYNAMICS AND DYSFUNCTION. bioRxiv. 2019; .
52. Bernard C, Jirsa V. Virtual brain for neurological disease modeling. Drug Discov Today Dis Model. 2016; 19:5–10.